\documentstyle[aps,epsfig,multicol,amssymb,amsfonts]{revtex}
\tighten
\renewcommand{\widetext}
{\end{multicols}\global\columnwidth42.5pc}
\begin{document}
\newcommand{\be}{\begin{equation}}
\newcommand{\ee}{\end{equation}}
\newcommand{\bea}{\begin{eqnarray}}
\newcommand{\eea}{\end{eqnarray}}
\newcommand{\br}{{\bf r}}
\newcommand{\bk}{{\bf k}}
\newcommand{\bq}{{\bf q}}
\newcommand{\bn}{{\bf n}}
\draft
\title{Cyclotron resonance in antidot arrays}
\author{D. G. Polyakov,$^{1,*}$ F. Evers,$^1$ and I. V. Gornyi$^{1,2,*}$}
\address{$^1$Institut
f\"ur Nanotechnologie, Forschungszentrum Karlsruhe, 76021 Karlsruhe,
Germany}
\address{$^2$Institut f\"ur Theorie der Kondensierten Materie,
Universit\"at Karlsruhe, 76128 Karlsruhe, Germany}
\maketitle
\begin{abstract}
We study the dynamical properties of an electron gas scattered by
impenetrable antidots in the presence of a strong magnetic field. We
find that the lineshape of the cyclotron resonance is very different
from the Lorentzian and is not characterized by the Drude scattering
rate. We show that the dissipative dynamical response of skipping
orbits, $S_c(\omega)$, is broadened on a scale of the cyclotron
frequency $\omega_c$ and has a sharp dip $\propto
|\omega-\omega_c|$. For small antidots, $S_c(\omega)$ is strongly
modulated with a period equal to $\omega_c$ and has sharp square-root
singularities for a series of resonant frequencies. For large
antidots, $S_c(\omega)$ has a hard gap at $\omega<\omega_c$ between
two sharp peaks, associated respectively with edge states and free
cyclotron orbits.
\end{abstract}
 
\vspace{0.5cm}
\begin{multicols}{2}

\narrowtext
\section{Introduction}
\label{sec1}
 
Progress in controlled fabrication of semiconductor nanostructures
\cite{ferry97} has revived interest in quasiclassical features of
transport in a two-dimensional electron gas (2DEG). Since the Fermi
wavelength of electrons in high-mobility heterostructures is usually
small compared to the characteristic spatial scale of inhomogeneities,
the transport properties of the 2DEG retain signatures of the
underlying quasiclassical dynamics of the particles. In particular,
transport in ballistic mesoscopic systems, where electrons are
scattered specularly on the boundary of the system, has been
investigated in terms of quasiclassical dynamics in considerable
detail \cite{beenakker91}. In antidot (AD) arrays, potential barriers
around the ADs can also be viewed as hard discs of size $\sim
10-10^2$~nm that reflect electrons specularly. If AD arrays are
periodic (for a review see \cite{weiss97,fleischmann95}), the
quasiclassical character of electron dynamics manifests itself in
pronounced geometrical resonances, which are associated with the
periodicity and lead, in particular, to commensurability peaks in the
magnetoresistance. On the other hand, random AD arrays (for
experimental work on dc transport see, e.g.,
\cite{gusev94,tsukagoshi95,nachtwei98,yevtushenko00}) represent a
remarkable disordered system in which the statistics of fluctuations
of the random potential is strongly non-Gaussian (in contrast to the
familiar case of smooth disorder in high-mobility heterostructures,
where the random potential at a given point is a sum of contributions
from many impurities).
 
On the theoretical side, the recent interest in the quasiclassical
dynamics of a 2DEG is to a large extent inspired by a variety of
``non-Boltzmann" quasiclassical transport phenomena that occur in
disordered systems with large-scale inhomogeneities. The term
``non-Boltzmann" means that these phenomena, while being essentially
classical, cannot be described by the Boltzmann kinetic equation
(i.e., the collision-integral approximation is insufficient). They are
due to correlations of scattering acts at the points where
quasiclassical paths self-intersect, which gives rise to memory
effects, not captured by Boltzmann transport theory. Most noticeably,
the non-Markovian kinetics yields a wealth of anomalous
magnetotransport phenomena in low magnetic fields
\cite{mirlin99,mirlin01,polyakov01}, induces adiabatic localization of
electrons in strong fields \cite{fogler97,evers99}, and leads to a
peculiar behavior of the magnetoresistivity in the Lorentz model of
hard-disc scatterers \cite{baskin78,bobylev95,kuzmany98}. It is
important that the quasiclassical non-Boltzmann corrections dominate
over the quantum ones in systems with long-range disorder.

The non-Markovian character of kinetics that leads to the anomalous dc
transport manifests itself also in the cyclotron resonance (CR). In
particular, the adiabatic localization \cite{fogler97,evers99} is
predicted \cite{fogler98} to yield a peculiar shape of the CR line: a
narrow peak related to chaotic dynamics of delocalized electrons on
top of an inhomogeneously broadened background coming from
adiabatically localized electrons. The latter contribution dominates
for large $B$ and gives the linewidth which depends nonmonotonically
on $B$. This intricate picture should be contrasted with the case of
white-noise disorder, where the linewidth is given by the scattering
rate \cite{ando75}.
 
In this paper, we consider the CR in AD arrays. We demonstrate that
the CR lineshape is very much different from the Lorentzian suggested
by the Drude theory. The peculiarity of the dynamical response of the
AD system is related to two factors which become important with
increasing $B$: formation of ``skipping orbits" of electrons bound to
ADs and suppression of scattering for other electrons that do not
participate in the process of skipping.  We show that the CR is not
characterized simply by the Drude scattering rate. Specifically,
cyclotron orbits not colliding with ADs yield an infinitely sharp CR
line, whereas skipping orbits give a contribution broadened on a scale
of the cyclotron frequency $\omega_c$. The skipping-orbit contribution
exhibits a remarkably rich behavior as a function of frequency
$\omega$: it has a sharp dip $\propto |\omega-\omega_c|$ at
$\omega=\omega_c$ and, in the case of small ADs, is strongly modulated
with a period equal to $\omega_c$. The modulation yields exact zeros
of the CR response in a dilute AD array. In addition to the zeros, a
series of sharp singularities is developed in the wings of the CR
line. For large ADs, the dynamical response has a hard gap at
$\omega<\omega_c$ between two sharp peaks, associated respectively
with edge states and free cyclotron orbits.

The paper is organized as follows. In Sec.~\ref{sec2}, we derive an
exact expression for the CR lineshape in the limit of large $B$, which
is then analyzed in two essentially different cases of small
(Sec.~\ref{sec3}) and large (Sec.~\ref{sec4}) ADs. In Sec.~\ref{sec5},
we consider moderately strong magnetic fields (moderately in the sense
that skipping orbits bound to different ADs can overlap). We add
remarks bearing on experiment and the role of electron-electron
interactions in Sec.~\ref{sec6}. Throughout the paper, the analytical
calculation is complemented by results of numerical simulations.

\section{Large-$B$ limit: Insulating phase} 
\label{sec2}

We start by considering the limit of large $B$, namely $n_SR_c^2\ll
1$, where $n_S$ is the concentration of ADs, $R_c$ the cyclotron
radius. We model ADs by hard discs and assume that the array of ADs is
dilute, i.e., $n_Sa^2\ll 1$, where $a$ is the radius of the discs. In
this model (known as the Lorentz gas), if $R_c$ is smaller than a
critical value of the order of $n_S^{-1/2}$, all quasiclassical
trajectories get localized and the dissipative dc conductivity
$\sigma_{xx}(\omega=0)$ vanishes at zero temperature exactly
\cite{baskin78,bobylev95}. The localization is developed through the
formation of disconnected clusters of trajectories that do not extend
beyond a finite area. In random AD arrays, the metal-insulator
transition is ``second order", so that as one goes deeper in the
insulating phase with increasing $B$, the size of the critical
clusters decreases continuously. Eventually, far away from the
critical point, for $n_SR_c^2\ll 1$, trajectories that collide with
two or more ADs become very rare. In this limit, most electrons do not
collide with ADs whatsoever and give a $\delta$-function CR line at
$\omega=\omega_c$. Most of those that collide move in skipping orbits
around a single AD. It follows that for large $B$ the dynamical
response at $\omega\neq\omega_c$ is determined by the skipping
orbits. Clearly, this conclusion is true both for random and periodic
AD arrays.
 
Let us calculate the dynamical response associated with skipping
orbits.  Since in phase space of the Lorentz gas there is a
well-defined separatrix between free cyclotron orbits and trajectories
colliding with ADs, we write the dissipative conductivity as a sum of
two terms,                             
\begin{equation}
{\rm Re}\,\sigma_{xx}(\omega)={1\over
2}e^2\rho_0[\,pD_f(\omega)+(1-p)D_c(\omega)]~,
\label{1}
\end{equation}
where
\begin{equation}
p=\exp (-2\pi/\omega_c\tau_0)
\label{2}
\end{equation}
is the probability to close the cyclotron orbit without suffering a
collision, $\tau_0=1/2v_Fn_Sa$ the collision time, $\rho_0$ the
density of states for free electrons at $B=0$. The symmetrized
functions $D_{f,c}(\omega)=v_F^2[S_{f,c}(\omega)+S_{f,c}(-\omega)]/2$
are the velocity-velocity correlators for free electrons and electrons
colliding with ADs, respectively; $v_F$ is the Fermi velocity.  In the
above, we have assumed that many Landau levels are occupied, so that
$k_FR_c\gg 1$, where $k_F$ is the Fermi wavevector, and that ADs are
large enough, in the sense that $k_Fa\gg 1$, which means that ADs
scatter electrons specularly. It is also important to us that under
these conditions the density of states of electrons scattered by ADs
is not affected by the Landau quantization and is given simply by
$(1-p)\rho_0$.
 
The dynamical response to a circularly polarized (CR-active)
perturbation is given by $S_{f,c}(\omega)$. For free electrons we have
$S_f(\omega)=\pi\delta(\omega-\omega_c)$, while for electrons skipping  
around an AD
\begin{eqnarray}
S_c(\omega)&=&\int^\infty_0 \!\! dt\, \langle\cos[(\omega_c-\omega)t
\nonumber  \\
 &+&2\theta\sum_{m=1}^{\infty}\Theta(t+\Delta t
- m T)] \rangle_{\Delta t,r}~,
\label{3}
\end{eqnarray}
where $\Theta (t)$ is the step function and
\begin{equation}
\left<\,\,\right>_{\Delta t}={\omega_c\over
2\pi}\int_0^{T(r)}\!\!\!\!d\Delta t~,\quad \left<\,\,\right>_r={1\over
2R_ca}\int_{|R_c-a|}^{R_c+a} \!\! rdr
\label{4}
\end{equation}
denotes averaging over the initial phase $\omega_c\Delta t$ and the
distance $r$ between the centers of the AD and the cyclotron orbit. The
normalization of the integral $\left<\,\right>_{\Delta t}$ takes care
of the ``exclusion volume" free of electrons because of the presence
of ADs. The angle of incidence $\theta(r)$ (the angle between the
trajectory and the tangent to the surface of the AD at the collision
point) and the
time $T(r)$ between two successive collisions read
\begin{eqnarray}
\cos\theta (r)&=&\frac{r^2-R_c^2-a^2}{2 R_c a}~, \label{5} \\
\cos{\omega_c T(r) \over 2}&=&\frac{a^2-R_c^2-r^2}{2 R_c r}~.
\label{6}
\end{eqnarray}
Since $r$ is an integral of motion for orbits skipping around a disc,
$\theta (r)$ and $T(r)$ are the same for each collision.
 
Doing integrals over $t$ and $\Delta t$ and summing over $m$ in
Eq.~(\ref{3}), we get an identically zero response at the cyclotron
frequency for colliding electrons, $S_c(\omega_c)\equiv 0$, and
\bea
S_c(\omega)&=&\frac{\omega_c}{(\omega_c-\omega)^2}
\left<\sin^2 \theta(r) \!\sum_{n=-\infty}^\infty\!
\delta[f_\omega(r)+\pi n]\right>_r
\label{7}\\
&=& \frac{\omega_c}{2R_c a
(\omega_c-\omega)^2}\sum_n
\frac{r_n\sin^2\theta(r_n)}{|f'_\omega(r_n)|}
\label{8}
\end{eqnarray}
for $\omega \neq \omega_c$. Here $f_\omega(r)={1\over
2}(\omega_c-\omega)T(r)+\theta (r)$ and $r_n(\omega)$ are roots of the
equation $f_\omega(r)+\pi n=0$. The summation in Eq.~(\ref{8}) runs
over $r_n$ that satisfy $|R_c-a|<r_n<R_c+a$.
 
Equation (\ref{7}) tells us that the dynamical response is due to
resonant orbits. There is no broadening of the contribution of each of
the orbits and, accordingly, the shape of the CR line is given by the
density of states of the resonant orbits. The meaning of the resonance
condition $\omega=\omega_c+2(\theta +\pi n)T^{-1}$ is that the change
of the total phase (cyclotron phase + phase of the ac field +
scattering phase) should be a multiple of $2\pi$ between two
collisions. It is worth noting that, contrary to what one might
expect, the resonant orbits are not periodic two-dimensional
orbits. While periodic orbits and trajectories close to them are
indeed important in periodic arrays at weak $B$, where they give rise
\cite{weiss91,fleischmann92} to geometric resonances, they do not play
any particular role in Eqs.~(\ref{7}),(\ref{8}). This conclusion
should be contrasted with the approach of Ref.~\cite{vasiliadou95},
where the ``periodic-orbit analysis" \cite{fleischmann92}, applicable
to the case of weak $B$, was extended to describe also the
experimentally observed dependence of the position of the CR line on
$B$ in an AD array for skipping orbits bound to a single AD. Our
calculation shows that this procedure is not justified.
 
To analyze $S_c(\omega)$, it is convenient to rewrite Eq.~(\ref{8}) and the
resonance condition in terms of the function
$g_\omega(\theta)=f_\omega[r(\theta)]$ defined on the interval
$0<\theta<\pi$:
\begin{eqnarray}
S_c(\omega)=\frac{\omega_c}{2(\omega_c-\omega)^2}\sum_n
\frac{\sin^3 \theta_n}{|g'_\omega(\theta_n)|}~;
\label{10} \\
g_\omega(\theta)={1\over 2}(\omega_c-\omega)T(\theta)+\theta~;
\label{10b} \\g_\omega (\theta_n)+\pi n=0~. \label{10a}
\end{eqnarray} 
The time between collisions $T(\theta)$ behaves in an essentially
different way depending on whether the ratio $R_c/a$ is larger or
smaller than unity.
 
\section{Small antidots}
\label{sec3}
 
Let us first study the case $R_c\gg a$ by expanding $T(\theta)$ in
powers of $a/R_c$. Combining Eqs.~(\ref{5}),(\ref{6}) we get
\begin{equation}
T(\theta)={2\pi\over \omega_c}-{2a\over v_F}\left( \sin\theta-{a\over
2R_c}\sin 2\theta+\dots \right)
\label{11}
\end{equation}
with $T(0)=T(\pi)\equiv 2\pi/\omega_c$.  To find the roots $\theta_n$
at $|\omega-\omega_c|\ll v_F/a$, one can retain only the first
(unperturbed) term in Eq.~(\ref{11}), which yields, for a given
$\omega$, a single solution $\theta_n\simeq\pi
(\omega/\omega_c-[\omega/\omega_c])$ and $g'_\omega(\theta_n)\simeq
1$. Here $[\omega/\omega_c]$ is the integer part of
$\omega/\omega_c$. Substituting these expressions in Eq.~(\ref{10}) we
find
\begin{eqnarray}
S_c(\omega)&=&\frac{\omega_c}{2(\omega_c-\omega)^2}
\,\sin^3\frac{\pi(\omega-n\omega_c)}{\omega_c}~,
\label{12} \\
D_c(\omega)&=&
v_F^2\frac{\omega_c^2+\omega^2}{(\omega_c+\omega)^2}   
S_c(\omega)
\label{13}
\end{eqnarray}
for $n\omega_c <\omega < (n+1)\omega_c$ and $|n|\ll R_c/a$.  We thus
see that the CR line exhibits a strong modulation, namely
$S_c(\omega)$ has zeros at $\omega=n\omega_c$. Note that the zeros are
exact even if higher-order terms in powers of $a/R_c$ are taken into
account.  These zeros correspond to resonant orbits with $\theta\to 0$
and $\theta\to \pi$, which go along the tangent at the point where
they touch the AD.  In the vicinity of the zeros $S_c(\omega)$ behaves
as $|\omega-n\omega_c|^3$ for any $n\neq 1$, including the dc limit
$n=0$. At the point $\omega=\omega_c$, $S_c(\omega)$ vanishes as
$|\omega -\omega_c|$. The envelope of the oscillations (\ref{12})
falls off with increasing $\omega$ as $(\omega-\omega_c)^{-2}$,
similarly to the conventional Lorentzian.
 
\begin{figure}
\begin{center}
\includegraphics[width=0.95\columnwidth,clip]{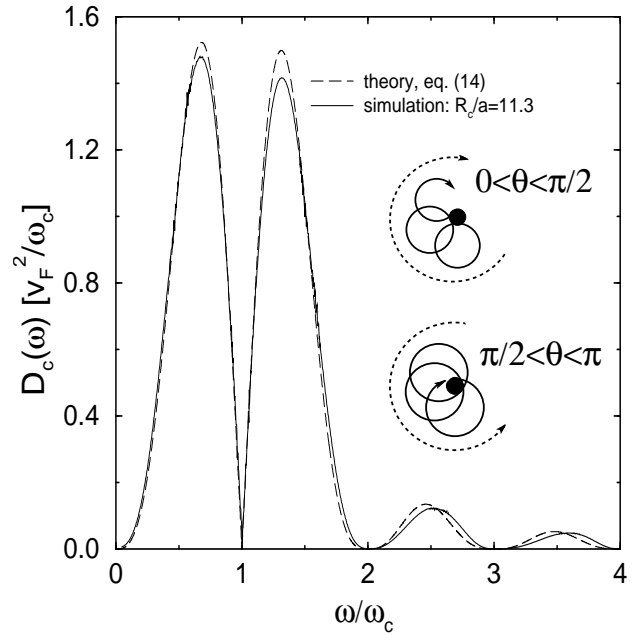}
\end{center}
\caption{Dynamical response of skipping orbits for $R_c/a\gg 1$ and
$\omega\ll v_F/a$ exhibits oscillatory behavior with a characteristic
double peak around zero at $\omega=\omega_c$. Dashed line:
$D_c(\omega)$ in units of $v_F^2/\omega_c$ according to
Eqs.~(\ref{12}),(\ref{13}); solid line: the numerical simulation for
$R_c/a\simeq 11.3$. The insets illustrate the different sense of
rotation for resonant skipping orbits with $1<\omega/\omega_c<3/2$
(upper panel) and
$1/2<\omega/\omega_c<1$ (lower panel).}
\label{fig1}
\end{figure}

The total response of the 2DEG scattered by ADs of small radius $a$
($R_c/a\gg 1$ and $|\omega|\ll v_F/a$) is thus a sum of two parts: a
sharp peak associated with $S_f(\omega)$ and a series of broader peaks
given by $S_c(\omega)$, whose width is $\sim\omega_c$. The oscillatory
behavior of $D_c(\omega)$ is shown in Fig.~\ref{fig1}. One can see
that the first two peaks of $D_c(\omega)$ that occur between
$\omega=0$ and $\omega=2\omega_c$ are much higher than those for
larger $\omega$. As a result, the response of skipping orbits with
$R_c/a\gg 1$ looks like a double peak split up at
$\omega=\omega_c$. Note that the resonant value of $\theta$ tends to
$\pi$ if one approaches $\omega_c$ from the left
($\omega\to\omega_c-0$), whereas it tends to 0 if one does so from the
right ($\omega\to\omega_c+0)$. Accordingly, the two parts of the
double peak for $1<\omega/\omega_c<3/2$ (where the resonant angles $0<\theta<\pi/2$) and
$1/2<\omega/\omega_c<1$ (where $\pi/2<\theta<\pi$) differ in the direction in which the guiding centers of the resonant
skipping orbits rotate around the AD (see the insets in Fig.~\ref{fig1}).
 
The wing of the principal double peak [Fig.~\ref{fig1}] exhibits
nontrivial behavior. As $\omega$ increases, $S_c(\omega)$ in
intervals between two adjacent zeros gets more and more asymmetric
with maxima shifting towards the higher-$|\omega|$ boundary of the
intervals. Eventually, when $|\omega|$ reaches the critical frequency
$v_F/a$, the behavior of $S_c(\omega)$ acquires qualitatively new  
features. At $|\omega|>v_F/a$, $g_\omega(\theta)$ as a function of
$\theta$ becomes nonmonotonic, which can be seen from
Eqs.~(\ref{10b}),(\ref{11}). This leads to the appearance of multiple
roots $\theta_n$ for a given $\omega$ (in contrast to a single root at
$|\omega|<v_F/a$).  Because of the multiple roots, the zeros in
$S_c(\omega)$ disappear for $|\omega|>v_F/a$, since the condition
$\theta_n=0$ or $\pi$ now cannot be met for all roots simultaneously
(so that there is a finite number of zeros, namely $2R_c/a$ zeros at
$R_c\gg a$). Moreover, the nonmonotonic behavior of $g_\omega(\theta)$
yields singularities in the lineshape associated with resonant orbits
for which the derivative $g'_\omega(\theta_n)$ vanishes in the
denominator of Eq.~(\ref{10}).

\begin{figure}
\begin{center}
\includegraphics[width=0.95\columnwidth,clip]{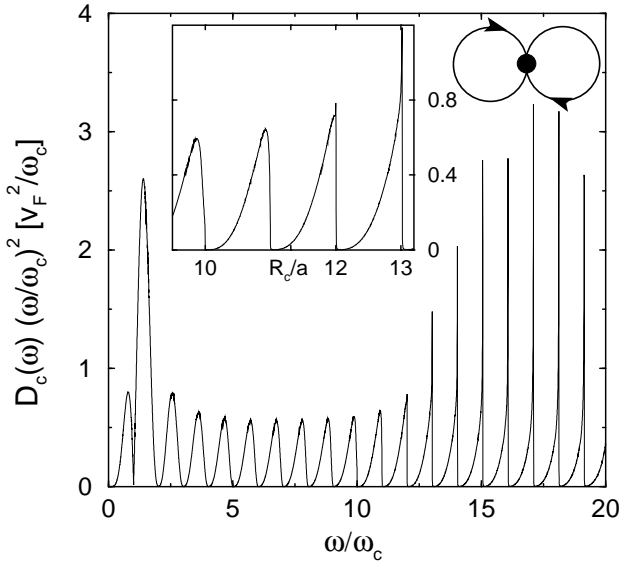}
\end{center}
\caption{Dynamical response of skipping
orbits with $R_c/a\gg 1$ in the tail of the principal double peak
(shown in Fig.~\ref{fig1}). The oscillating curve
represents the product $D_c(\omega)\times (\omega/\omega_c)^2$ in units of
$v_F^2/\omega_c$
as obtained from the numerical simulation for $R_c/a\simeq
11.3$. The inset magnifies the region near $\omega/\omega_c =R_c/a$,
where zeros of $D_c(\omega)$ disappear and
simultaneously a series of square-root singularities starts
[Eqs.~(\ref{14}),(\ref{15})]. It also shows the resonant orbit
corresponding to the singular frequency. }
\label{fig2}
\end{figure}     

Let us analyze the high-$\omega$ limit $|\omega|\gg v_F/a$.  In this
case, one can neglect the last term in the right-hand side of
Eq.~(\ref{10b}) and represent the equation for $\theta_n$ at $n\gg 1$
in the form $(\omega a/v_F)\sin \theta_n+\pi (n-\omega/\omega_c)=0$
with $g'_\omega(\theta_n)\simeq (\omega a/v_F)\cos\theta_n$. One sees
that now for a given $\omega$ there are $2 |\omega| a/\pi v_F\gg 1$
roots $\theta_n$. Transforming to the continuous limit in the
summation over $\theta_n$ [Eq.~(\ref{10})] we obtain a regular
part of $S_c(\omega)$ for $|\omega|\gg v_F/a$:
\begin{equation}
S_c^{reg}(\omega)={2\over 3\pi}\,{\omega_c\over \omega^2}~.
\label{14}
\end{equation}    
The most prominent feature on top of the smoothly varying background
(\ref{14}) is the appearance at $|\omega|>v_F/a$ of sharp singularities
(``spikes") in $S_c(\omega)$. The spikes occur every time
$g'_\omega(\theta)$ vanishes for one of the roots $\theta_n$. Note
that the singularity frequencies coincide for $S_c(\omega)$ and
$S_c(-\omega)$. At $|\omega|\gg v_F/a$, $g_\omega(\theta)$ reaches
maximum at $\theta=\theta_{max}\simeq \pi/2+a/R_c+v_F/\omega a$, close
to $\pi/2$.  It follows that the skipping orbits that yield the
spikes hit the surface of an AD at almost a right angle, so that the
center of the orbits flicks from one side of the AD to the other after
each collision. Expanding around the maximum and substituting two
(almost degenerate) roots $\theta_n(\omega)$ of Eq.~(\ref{10a}) which
are close to $\theta_{max}$, we find $S_c(\omega)$ in the vicinity of
the singularities for $|\omega|\gg v_F/a$:
\begin{eqnarray}
S_c^{sing}(\omega)&=& {\omega_c\over\omega^2}\left({v_F\over 2\pi
a|\omega|}\right)^{1/2}\nonumber \\ &\times& \left|{\omega_c\over
\omega_n-\omega}\right|^{1/2}\Theta[(\omega_n-\omega)\,\,{\rm sgn}\,
\omega]~.
\label{15}
\end{eqnarray}
The frequencies $\omega_n$ for $|\omega|\gg v_F/a$ and $R_c\gg a$ are
given by $\omega_n\simeq \omega_c(n+3/2)(1+a/\pi R_c)$. Note that the
period of the sequence of spikes is larger than $\omega_c$ (but
close to $\omega_c$ at $R_c\gg a$).
The behavior of $S_c(\omega)$ for large $\omega a/v_F$ is
illustrated in Fig.~\ref{fig2}. The square-root singularities at   
$\omega\to\omega_n$ that appear for  $\omega> v_F/a$  (i.e., for
$\omega/\omega_c >R_c/a$) are clearly seen. The region of
$\omega\simeq v_F/a$ is blown up in the inset to show the sudden
start of the series of
singularities when $\omega$ becomes larger than $v_F/a$.
 
\section{Large antidots} 
\label{sec4}

We now turn to the case $R_c\ll a$. As we will see, $S_c(\omega)$
shows completely different behavior in the two limits of large and
small $R_c/a$. The function $T(\theta)$ expanded in powers of $R_c/a$
reads:
\begin{equation}
T(\theta)={2(\pi -\theta)\over \omega_c}+{2R_c\over \omega_c
a}\left(\sin\theta-{3R_c\over 4a}\sin 2\theta +\dots\right)
\label{18}
\end{equation}
with $T(0)\equiv 2\pi/\omega_c$ and $T(\pi)\equiv 0$. Clearly, if one
sends $R_c/a\to 0$, $S_c(\omega)$ describes the dynamical response of
trajectories skipping along a straight line (``edge states"). Using
(\ref{18}) in this limit, we get $1+[|\omega|/\omega_c]$ roots
$\theta_n=\pi-(n+1)\pi\omega_c/|\omega|$ with $0\leq n+1\leq
[|\omega|/\omega_c]$, for which
$g'_\omega(\theta_n)=\omega/\omega_c$. Substitution in Eq.~(\ref{10})
yields
\begin{equation}
S_c(\omega)={\omega_c^2\over
2|\omega|(\omega_c-\omega)^2}\sum_{n=0}^{[|\omega|/\omega_c]-1}
\sin^3{(n+1)\pi\omega_c\over |\omega|}
\label{19}
\end{equation}
and $D_c(\omega)$ which is related to $S_c(\omega)$ by Eq.~(\ref{13}).
Since the limits $\omega\to 0$ and $R_c/a\to 0$ do not commute with  
each other, one should be careful about the behavior of $S_c(\omega)$
at $\omega\to 0$. In fact, $S_c(\omega)$ for the edge states has a
$\delta (\omega)$-term in the dc limit, so that
Eq.~(\ref{19}) correctly describes the response of a large AD with
$R_c/a\to 0$ only at $\omega\neq 0$ (see below).
 
As follows from Eq.~(\ref{19}), in a striking difference with the case
of large $R_c/a$, there appears a gap with $S_c(\omega)=0$ for
$0<|\omega|<\omega_c$. No dissipation occurs with increasing $\omega$
until $|\omega|$ exceeds $\omega_c$. At the edge of the gap,
$S_c(\omega)$ vanishes linearly as $|\omega|-\omega_c$. Also, in
contrast to the case of large $R_c/a$, $S_c(\omega)$ given by
Eq.~(\ref{19}) has neither zeros [cf.\ Eq.~(\ref{12})] nor
singularities [cf.\ Eq.~(\ref{15})] for $|\omega|>\omega_c$. In the
limit of large $|\omega|\gg \omega_c$, replacing the summation in
Eq.~(\ref{19}) by integration leads to $S_c(\omega)\propto
\omega^{-2}$ which behaves according to Eq.~(\ref{14}).
 
\begin{figure}
\begin{center}
\includegraphics[width=0.95\columnwidth,clip]{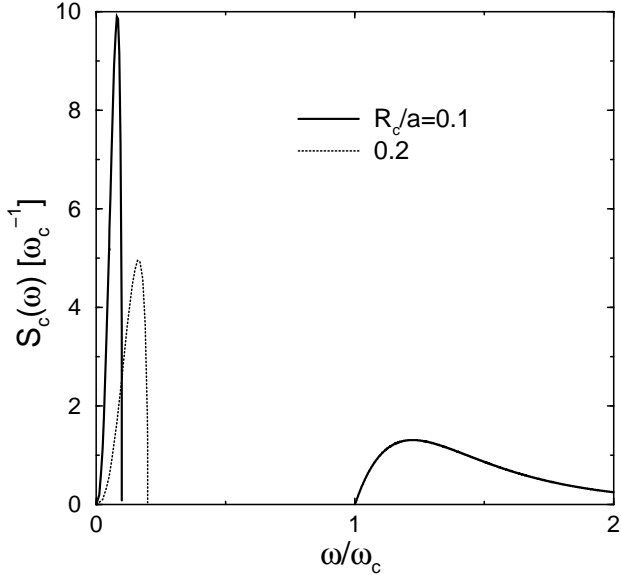}
\end{center}
\caption{Dynamical response of skipping orbits with $R_c/a\ll 1$
(edge states). The sharp low-frequency peak is separated from the
broad peak at $\omega\sim \omega_c$ by a hard gap. The curve for
$\omega>\omega_c$ is $S_c(\omega)$ in units of $\omega_c^{-1}$
calculated according to Eq.~(\ref{19}). The peak at $0<\omega< v_F/a$
is a solution of Eqs.~(\ref{20}),(\ref{21}) for $R_c/a=0.1$ (solid
line) and $R_c/a=0.2$ (dotted line).}
\label{fig3}
\end{figure}
 
As mentioned above, in addition to Eq.~(\ref{19}) there is a $\delta
(\omega)$ peak in $S_c(\omega)$ for edge states with $R_c/a\to 0$. To
calculate the dynamical response of ADs of a large but finite radius,
one should take into account terms of higher order in $R_c/a\ll 1$ in
the expansion (\ref{18}), which leads to a broadening of the peak in a
finite range of frequency, namely $0<\omega<v_F/a$. This broadening is
governed by the root $\theta_{-1}$ which obeys the equation  
\begin{equation}
{\omega
a\over v_F}={\sin\theta_{-1}\over \pi-\theta_{-1}}
\label{20}
\end{equation}
with $g'_\omega
(\theta_{-1})=\omega_c^{-1}(\omega-v_F\cos\theta_{-1}/a)$, where we
retained only the term linear in $R_c/a$ in Eq.~(\ref{18}). We see
that a solution of (\ref{20}) exists only in the abovementioned
interval of $\omega$. Within this interval $S_c(\omega)$ reads
\begin{equation}
S_c(\omega)={a\over 2v_F}\,{\sin^3\theta_{-1}\over \cos\theta_{-1}+
\omega a/v_F}~.
\label{21}
\end{equation}
Combining Eqs.~(\ref{20}),(\ref{21}) we obtain the asymptotic behavior
of $S_c(\omega)$ near the ends of the interval: $S_c(\omega)$ vanishes
as $\omega^3$ at $\omega\to 0$ and as $(v_F/a-\omega)^{1/2}$ at
$\omega\to v_F/a$. The total weight of the peak at small $\omega$ is
given by $\int_0^{v_F/a}\!\!d\omega\, S_c(\omega)=\int_0^\pi\!
d\theta\, (2\theta)^{-1}\sin^3\theta \simeq 0.485$.
 
The dynamical response of the 2DEG in the case of relatively large ADs
with $R_c/a\ll 1$ is thus a sum of three peaks: two sharp peaks, one
centered at low frequency $\omega\sim v_F/a$, the other at higher
frequency $\omega=\omega_c$ (the latter is related to free cyclotron
orbits), plus a broad peak of width $\sim\omega_c$ whose edge
coincides with $\omega_c$. The behavior of the contribution of
skipping orbits, $S_c(\omega)$, is illustrated in Fig.~\ref{fig3}.
 
Note that the hard gap in $S_c(\omega)$ gets narrower but
remains exact at small $R_c/a$. In fact, the hard gap survives
with increasing $R/a$ up until $R_c$ becomes equal to $a$, at which
point the geometry of skipping changes in a qualitative way.
Specifically, at $R_c<a$ skipping orbits propagate in only one
direction around the AD, whereas for $R_c>a$ they propagate in both.
As a result, the hard gap at $|\omega|<\omega_c$ disappears 
(transforms into a soft gap $\propto |\omega-\omega_c|$) for $R_c>a$
and simultaneously infinitely many singularities pop up at
$|\omega|>\omega_c$. With further increasing $R_c/a$, zeros in
$S_c(\omega)$ proliferate in the finite range of frequency
$|\omega|<v_F/a$, as explained above.

\section{Moderately strong $B$: Metallic phase}  
\label{sec5}

The above analysis of the dynamical response of a single AD applies
directly to the case of a strong magnetic field, $n_SR_c^2\ll 1$. In
this limit, a skipping orbit is bound to a single AD for an infinitely
long time. With decreasing $B$, the scattering processes that involve
collisions of a skipping orbit with many ADs become essential and at a
critical value of the parameter $n_SR_c^2\sim 1$ a metal-insulator
transition occurs. On the conducting side of the transition, the
skipping orbits become delocalized by hopping from one AD to
another. Deep in the metallic phase, for $n_SR_c^2\gg 1$, the
characteristic hopping rate is $\tau_0^{-1}$ and the dc conductivity
$\sigma_{xx}(0)$ is given by the second term in Eq.~(\ref{1}) with
\cite{bobylev95} $D_c(0)=v_F^2\tau(x)/[1+\omega_c^2\tau^2(x)]$
parametrized by $x=\omega_c\tau_0$, where $\tau (\infty)=\tau_0$ and
$\tau(0)=3\tau_0/4$. At zero $B$, Boltzmann theory works perfectly
well in the hydrodynamic limit ($n_S\to\infty$, $n_Sa={\rm const}$);
but, due to the factor of $1-p$ in Eq.~(\ref{1}), $\sigma_{xx}(0)$
falls off as $B^{-3}$ in the conducting phase, one power of $B$ faster
than in Boltzmann theory. One might think that, apart from this
factor, the dynamical response at finite $B$ will also be similar to
that in the Boltzmann approach. The latter is simply the zero-$B$
Lorentzian with a shifted frequency $\omega\to\omega-\omega_c$. In
fact, however, the behavior of $S_c(\omega)$ in the conducting phase
is completely different from the Lorentzian, see below.
 
While $S_c(\omega)$ in the dc limit shows the metal-insulator
transition, at larger $\omega$ the function $S_c(\omega)$ in the
conducting phase retains, provided $\omega_c\tau_0\gg 1$, the main
features of the single-AD response with $R_c/a\gg 1$
[Eqs.~(\ref{11})-(\ref{15})]. The oscillatory behavior of the
lineshape with sharp dips at $\omega=n\omega_c$ remains almost
unchanged at $\omega_c\tau_0\gg 1$, since in this limit skipping
orbits experience many collisions with a single AD before they
``change" to another one.  Clearly, in contrast to Eq.~(\ref{12}),
$S_c(\omega)$ has no exact zeros any more; however, the behavior of
$S_c(\omega)$ is modified only in a close vicinity of the points
$\omega=n\omega_c$. In particular, because of the broadening of the
resonances due to the hopping between ADs, the linear vanishing of
$S_c(\omega)\propto |\omega-\omega_c|$ near $\omega=\omega_c$ is cut
off at $|\omega-\omega_c|\sim \tau_0^{-1}$.  On the other hand, the
cubic vanishing of $S_c(\omega)\sim |\omega|^3/\omega_c^4$ matches
$S_c(0)\sim (\omega_c^2\tau_0)^{-1}$ in the dc limit, which
establishes the scale $|\omega|\sim \tau_0^{-1}(\omega_c\tau_0)^{2/3}$
on which the frequency dispersion of the conductivity becomes
strong. A linear zero-frequency anomaly \cite{wilke00}
$S_c(\omega)-S_c(0)\propto |\omega|$ appears in the metallic
phase. Notice that, because of the anomalously strong broadening of
the CR line, the crossover to the regime
$\sigma_{xx}(\omega)\gg\sigma_{xx}(0)$ occurs with increasing $\omega$
at much smaller (for $\omega_c\tau_0\gg 1$) frequency than in the
Drude regime, where the characteristic scale is $\omega_c$.

The hopping between ADs also cuts off the square-root singularities
(\ref{15}) [they only survive for isolated ADs, whose contribution is
suppressed by the factor of $\exp (-4\pi n_SR_c^2)$]. Note that the
substitution $\omega\to \omega+i/\tau_0$ in Eq. (16) is only correct
for $v_F/a \ll \omega \ll v_FR_c/a^2$. For larger $\omega$, the
effective collision rate for the resonant orbits is renormalized,
$\tau_0^{-1}\to \tilde{\tau}_0^{-1}$, since after each two collisions with a
given AD the center of the resonant orbit is shifted by a distance
$\delta R \sim R_c v_F/a\omega$, which is much smaller than $a$ for
the large frequencies.  It follows
that $\tilde{\tau}_0$ increases with $\omega$: $\tilde{\tau}_0/\tau_0\sim
a/\delta R$, so that the singularities are cut off on the smaller scale of
$\tilde\tau_0^{-1}$. Note that after the collision with another AD the
center of the orbit is shifted by $\sim R_c(\delta R/a)^{1/2}$, which
makes the orbit non-resonant.

The smearing of the soft gaps around the zeros of $S_c(\omega)$ can be
clearly seen in the hydrodynamic limit $n_S\to \infty$, $a\to 0$ with
$n_Sa$ held fixed. In this limit, the kinetic problem of finding
$\sigma_{xx}(\omega)$ allows for an exact solution
\cite{bobylev95,kuzmany98}, which we reproduce for convenience in the
following form:
\begin{eqnarray}
S_c(\omega)&=&{1\over \omega_c-\omega}\, \nonumber \\ &\times&{\rm
Re}\left[\,{1-pe^{-2\pi i\omega/ \omega_c}\over 1-p}\,\,{i\Sigma
(\omega)\over \omega-\omega_c-\Sigma(\omega)}\,\right]
\label{22}
\end{eqnarray}
with
\begin{equation}
\Sigma(\omega)={1\over \tau_0}\int_0^\pi\!\! d\theta
\,\sin^2\theta\, {e^{i\theta} \over 1-pe^{2ig_\omega(\theta)}}~.
\label{23}
\end{equation}
Here $g_\omega(\theta)$ is given by Eq.~(\ref{10b}) with
$T(\theta)=2\pi/\omega_c$, since in the hydrodynamic limit $R_c/a\to
\infty$.  Sending $\omega_c\tau_0\to\infty$ (i.e., $p\to 1$) generates
a pole on the real axis of $\theta$ in the integrand of the
self-energy (\ref{23}), which yields $\Sigma
(\omega)=(\pi/2\tau_0)\sin^2\theta e^{i\theta}$ with
$\theta=\pi(\omega/\omega_c-[\omega/\omega_c])$. The latter are
precisely the $\theta$'s defining the resonant orbits in
Eq.~(\ref{12}). Using the above expression for $\Sigma(\omega)$ at
$\omega_c\tau_0\to\infty$ in Eq.~(\ref{22}) indeed yields our
Eq.~(\ref{12}), derived for the insulating phase. Clearly, no
singularities [Eq.~(\ref{15})] occur in the hydrodynamic limit.  Note
that the metal-insulator transition takes place in a dilute AD array
at $R_c/a\gg 1$. Therefore, provided $\omega_c\tau_0\gg 1$ (which is
the conventional condition for a developed CR resonance),
$S_c(\omega)$ is in fact described by Eq.~(\ref{12}) not only in the
metallic phase but also in the critical region of the metal-insulator
transition, except for a very close vicinity of the zeros of
$S_c(\omega)$.

\begin{figure}
\begin{center}
\includegraphics[width=0.95\columnwidth,clip]{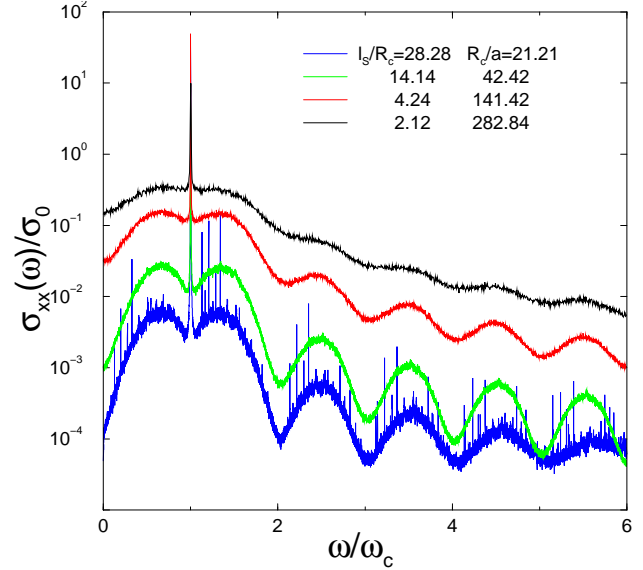}
\end{center}
\vspace{-3mm}
\caption{Dynamical conductivity $\sigma_{xx}(\omega)$ as a
function of $\omega/\omega_c$ in units of the zero-$B$ zero-$\omega$
value $\sigma_0$ in a dilute antidot array as obtained by the
numerical simulation. The transport mean free path at zero $B$ is
$l_S\simeq 600 a$. Different curves correspond to different $B$. The
ratio of $l_S/R_c$ changes from 2.12 for the upper curve to 4.24, to
14.14, up to 28.28 for the lower curve. Even for the relatively weak
field with $l_S/R_c=2.12$ a very sharp peak at $\omega=\omega_c$ is
clearly seen. With increasing $B$, the broad double peak around
$\omega=\omega_c$ gets more pronounced, as does the oscillatory
behavior of $\sigma_{xx}(\omega)$ for higher $\omega$. For large
$l_S/R_c=28.28$, the numerically obtained $\sigma_{xx}(\omega)$ agrees
well with Eqs.~(\ref{1}),(\ref{12}),(\ref{13}).  }
\label{fig4}
\end{figure}

In Fig.~\ref{fig4}, we show results of our numerical simulation of ac
transport in a very dilute AD array with $n_Sa^2\simeq 0.62\cdot
10^{-3}$, which corresponds to the mean free path $l_S=3/8n_Sa\simeq
600 a$, tuned through the metal-insulator transition by changing
magnetic field. The upper curve corresponds to the smallest $B$, with
$l_S/R_c\simeq 2.12$. One sees that, contrary to what one would expect
from the conventional Boltzmann theory, although the dc value
$\sigma_{xx}(0)$ is only $\sim 6$ times smaller than that at zero $B$,
a very narrow peak at $\omega=\omega_c$ is developed, associated with
free cyclotron orbits. As $B$ increases, this peak acquires more
weight, while remaining very sharp. Next, in striking contrast to
Boltzmann theory, the oscillations of $\sigma_{xx}(\omega)$, related
to skipping orbits, become more and more pronounced with growing
$B$. The double-peak structure around the linear gap near
$\omega=\omega_c$ is clearly seen, as are deep minima at $\omega$
equal to multiples of $\omega_c$. For large $l_S/R_c$, the numerically
obtained behavior of $\sigma_{xx}(\omega)$ is in a good agreement with
Eqs.~(\ref{1}),(\ref{12}),(\ref{13}).
 
As follows from the comparison of
Eqs.~(\ref{12}),(\ref{22}),(\ref{23}), the dissipation at $\omega\neq
\omega_c$ for finite $\omega_c\tau_0$ is due to resonant orbits
(\ref{10a}) broadened by the hopping between different ADs. It is
interesting to note that there appear two kinds of resonant behavior
of the dynamical response in the metallic phase: first, the last
factor in Eq.~(\ref{22}) resembles a resonance at
$\omega\simeq\omega_c$ broadened by ${\rm Im}\,\Sigma$; on the other
hand, the broadening itself is due to resonant orbits, for which the
denominator in Eq.~(\ref{23}) is close to zero.
 
\section{Discussion} 
\label{sec6}

In the above, we have calculated the quasiclassical dynamical
conductivity of an AD array, completely neglecting electron-electron
interactions inside the 2DEG. At zero $B$, in the case of Coulomb
interaction in a normal metal, this is a well-controlled approximation
governed by the large parameters $k_Fa_B$ and $k_F l_S$, where $a_B$
is the Bohr radius (and a characteristic screening length at zero
$B$). At nonzero $B$, the situation appears to be much less
trivial. Although in this paper we do not provide any treatment of the
combined effect of the interactions and disorder on the CR line, a few
comments are in order.
 
Firstly, the Landau quantization enhances the role of the
interaction. In particular, for weak $B$, the ground state of a
disorder-free 2DEG is known \cite{fogler96,moessner96} to
spontaneously break translational symmetry in a partially filled
(i.e. highest occupied) Landau level. It is electrons from this level
and those from the highest fully occupied level that are excited in
the CR at zero temperature. On the other hand, the CR at zero
wavevector in a one-component system of particles with a parabolic
dispersion without external inhomogeneities is insensitive to the
electron-electron interaction (Kohn's theorem \cite{kohn61}). Hence,
it is only because of the combined effect of disorder and interactions
that the latter can affect the CR lineshape (e.g.,
\cite{fukuyama79,kallin85} and references therein). At present, it is
unclear what is the resulting lineshape of the dynamical response for
the system studied in \cite{fogler96,moessner96} (see also
\cite{fradkin99,macdonald00,oppen00}) in the presence of
disorder. However, since electrons skipping around hard-wall ADs do
not experience the Landau quantization (in this sense disorder is
strong for them), we expect that their dynamical response has only
weak interaction corrections (for $k_Fa_B\gg 1$ and $k_FR_c\gg 1$). On
the experimental side, the CR of strongly correlated particles has
been investigated in the high-$B$ limit (e.g.,
\cite{wilson81,chou88,besson92,summers93} and references therein), but
not for the translational-symmetry broken state
\cite{fogler96,moessner96} in weaker magnetic fields.

Secondly, because of electron-electron interactions, the quantity that
is directly measured in far-infrared (e.g., \cite{batke86}) or
microwave (e.g., \cite{engel93,hohls01,ye01}) experiments may be
related to the conductivity in a complex way. In the paper, we have
provided explicit results for the conductivity; more specifically, for
$S_c(\omega)\propto {\rm Re}\,\sigma_{xx}(\omega)+{\rm Im}\,
\sigma_{xy}(\omega)$. We expect only small interaction corrections to
$S_c(\omega)$. It is important, however, that the conductivity
expresses the current as a response to the total (screened) electric
field. As such, it is given by the irreducible, with respect to the
Coulomb interaction, density-density response function $K({\bf q},{\bf
q}',\omega)$. On the other hand, in contrast to dc measurements, what
is probed directly in the ac transmission experiments is a response to
the external field. The latter is given, in an operator form, by the
reducible polarization $\epsilon^{-1}K $, where $\epsilon$ is the
dielectric function. Put differently, the dissipated power is measured
in units of the intensity of the incident wave. Since the measured
absorption is increased near zeros of $\epsilon$, i.e., on resonance
with plasma oscillations, the dynamical response of a 2DEG calculated
in the paper is in general masked in the absorption experiments by the
excitation of magnetoplasmons (for optical experiments with
magnetoplasmons see
\cite{batke86,kern91,zhao92,lorke92,bollweg95,vasiliadou95,bollweg96,hochgraefe99,cina99}). The
measured quantity appears to depend in an essential way on the
experimental setup (which may be very different, cf. \cite{hohls01}
and \cite{engel93,ye01}).

Let us mention one more point related to edge magnetoplasmons in AD
arrays, i.e., a soft mode of plasma oscillations localized near the
sharp edges of ADs (\cite{fessatidis93,mikhailov95}, for experiments
see
\cite{kern91,zhao92,lorke92,bollweg95,vasiliadou95,bollweg96,hochgraefe99,cina99}). It
is often asserted that there is an intimate connection between the
collective edge excitations and skipping orbits. In fact, this notion
may be very misleading. Indeed, the width of the strip around an AD in
which the skipping orbits propagate along the edge is $2R_c$. However,
the current in the edge magnetoplasmon mode decays, in general, on a
different scale: e.g., at $\omega\ll \omega_c$ and $R_c\gg a_B$ this
scale is given \cite{mikhailov95} by $R_c^2/a_B\gg R_c$. Hence, the
main contribution to the edge magnetoplasmon current may come from
electrons that do not at all collide with the AD. Since the frequency
dependence of the dissipative response of the skipping orbits and that
of the edge magnetoplasmons belong to different ranges of $\omega$, it
appears to be quite possible that they can be measured separately in
transmission experiments.

In conclusion, we have seen that the dynamical conductivity of
electrons scattered on impenetrable ADs in the presence of a magnetic
field reveals strong memory effects in the electron dynamics,
associated with skipping orbits bound to ADs. These lead to the CR
lineshape which is not at all characterized by the Drude scattering
rate. The contribution of the skipping orbits $S_c(\omega)$ is
broadened on a scale of the cyclotron frequency $\omega_c$ and
vanishes at $\omega_c$ in a nonanalytical way as
$|\omega-\omega_c|$. Apart from these two features, $S_c(\omega)$
exhibits different behavior depending on the ratio of the cyclotron
radius $R_c$ and the AD radius $a$. At large $R_c/a$, $S_c(\omega)$
oscillates with a period $\omega_c$ up to $\omega=\omega_cR_c/a$ and
shows a series of square-root spikes for larger $\omega$. At small
$R_c/a$, $S_c(\omega)$ has a hard gap between two sharp peaks located
at $\omega\sim \omega_c R_c/a$ and $\omega=\omega_c$.  We hope that
these results will stimulate further experimental work on the ac
conductivity in AD arrays.

\acknowledgements{We thank L.W. Engel, R. Haug, F. Hohls,
I.V. Kukushkin, U. Merkt, R.J. Nicholas, and J.H. Smet for discussions
concerning the experiments. This work was supported by SFB 195 and the
Schwerpunktprogramm ``Quanten-Hall-Systeme" of the Deutsche
Forschungsgemeinschaft, by INTAS Grants No.\ 99-1070 and 99-1705, and
by RFBR Grants No.\ 99-02-17093 and 00-02-17002.}

\end{multicols}
 
\end{document}